\documentclass[
nofootinbib,
noeprint,
amsmath,amssymb,
aps,
nobibnotes,
twocolumn,
floatfix,
]{revtex4-1}
\usepackage[utf8]{inputenc}
\usepackage{graphicx}
\usepackage{float}
\usepackage{epstopdf}
\usepackage{braket}

\usepackage{amsmath}
\usepackage{amssymb}
\usepackage{bm}
\usepackage{txfonts}

\usepackage{multirow}
\usepackage{mathptmx} 
\usepackage{siunitx}
\usepackage{caption}
\usepackage{subcaption}
\usepackage{xargs}                    
\usepackage[pdftex,dvipsnames]{xcolor}
\usepackage{datetime}
\ddmmyyyydate

\usepackage{tikz}
\usetikzlibrary{arrows.meta}
\setcitestyle{super}

\usepackage[colorlinks=true, urlcolor=blue, linkcolor=blue, citecolor=blue]{hyperref}

\begin{document}
\renewcommand{\thefootnote}{\roman{footnote}}

\title{Singular Azimuthally Propagating Electromagnetic Fields}
\author{Mustafa S. Bakr$^*$}\footnote[0]{$^\ast$These authors co-first author. \\ Email: mustafa.bakr@physics.ox.ac.uk, {smain.amari@physics.ox.ac.uk~}}
\author{Smain Amari$^*$} 
\normalsize{Department of Physics, University of Oxford, Clarendon Laboratory, Parks Road, Oxford OX1 3PU, U.K.}\hfill\\
\date{\today}

\begin{abstract}
We study the characteristics of azimuthally propagating electromagnetic fields in a cylindrical cavity. It is found that under certain conditions, the transverse components of the electromagnetic field are singular at the center of the cavity but the corresponding electromagnetic field remains of finite energy. The solutions are arranged in branches each of which starts from a root of $J_1(x)=0$ for the TE modes and a root of $J_0(x)=0$ for the TM modes. The lowest (dominant) branch starts from  a resonance that corresponds to the solution $x=0$ of $J_1(x)=0$. Its energy has a logarithmic singularity in a lossless structure. The singular solutions with finite energy can be observed experimentally by forcing them to resonate in a cavity with inserted metallic wedges. They can also be excited by transient sources. The singular electromagnetic field of these waves is strong enough to ionize the air. Whether these transient singular fields can initiate lightning, a phenomenon that is still not understood, is a very interesting question. It is also worth investigating whether the lowest resonance is excited in violently energetic cosmological phenomena such as cosmic jets.

\end{abstract}

\maketitle
Cylindrical cavities are found in various applications, such as microwave cavity filters\cite{1124039}, particle accelerators\cite{Yuan_2019}, and more recently in superconducting circuits for quantum computing\cite{MA20211789}. Traditionally, these cavities are characterised by the modes that can propagate in the axial direction with the $TE_{11}$ being the mode with the lowest cut-off frequency. The electromagnetic field inside the cavity is then modeled by the resonances that are brought about by this fundamental mode. However, models that are based on resonances are known to be less accurate than those based on propagating solutions, especially for wider bandwidth applications. Indeed, models based on propagation in the axial direction have been shown to provide efficient design techniques in microwave filtering\cite{4752854}. 

The work presented in this article was initiated in order to develop models that are based on solutions that are propagating in the azimuthal, instead of the axial, direction \cite{8563323}. Since a homogeneous cavity is invariant under a rotation in the azimuthal direction, or equivalently, around its axis, it supports simple propagating solutions.  Fortunately, the computation of these propagating solutions in the case of a cylindrical cavity, or a uniform section of a cylindrical cavity, is straightforward. However, the characteristics of the electromagnetic fields of some of these solutions turn out to be extremely interesting and surprising. Indeed, for small values of the propagation constant in the azimuthal direction, the transverse components of the electromagnetic field are singular at the center of the cavity. Even more surprising is the fact that propagation in the azimuthal direction starts from a resonance whose energy density varies with the radial distance $\rho$ as $\rho ^{-2}$ in the vicinity of the center of the cavity $\rho=0$. 
Although these singular solutions are not important in RF and microwave engineering, which was the original goal of this study, one cannot dismiss the possibility that they are behind some of the most violent natural phenomena. It is, for example, known that the electric field present between charged layers in thunderstorms is not strong enough to initiate lightening by ionising the air\cite{BUITINK20101}. However, the singularity of these azimuthally propagating electromagnetic waves, are certainly strong enough to ionize the air. Whether they can be excited by transient phenomena involving movement of electric charges in specific patterns for short time intervals as is the case in the atmosphere is a very interesting question which is not addressed in this paper. 
 In the next section, we examine azimuthally propagating electromagnetic fields in a cylindrical cavity. Since the structure is 'translationally' invariant in the azimuthal direction, it supports running solutions of the form $e^{\pm \nu i\phi}$. The first goal of the study is to determine how the propagation constant $\nu$ depends on frequency, or equivalently the dispersion relation of the local azimuthal waves.
\section{Transverse Electric modes}
 \begin{figure}[!h]
 {\includegraphics[width=1\columnwidth]{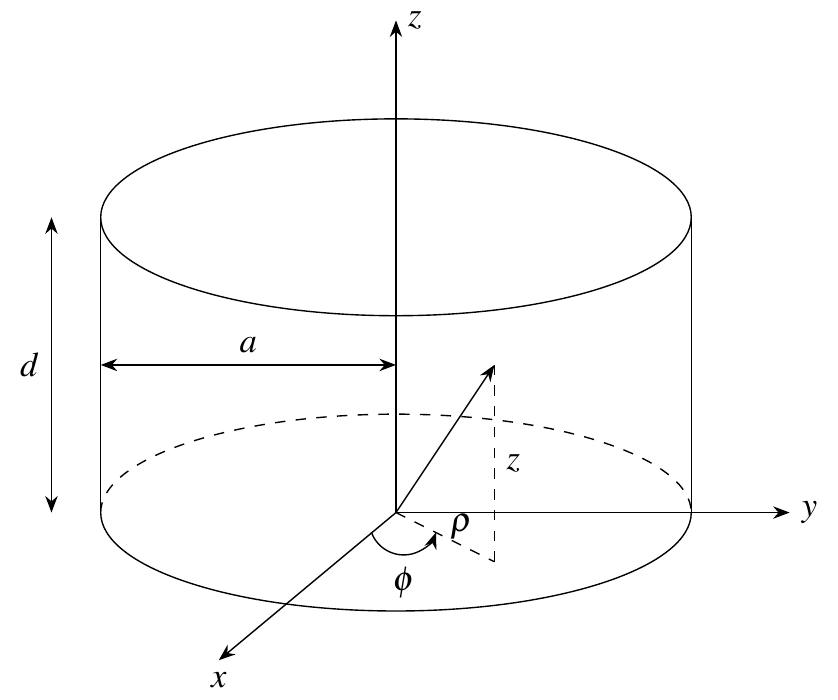}}
 \centering
 \caption{\label{wedge} 3D view of the cylindrical cavity.}
 \end{figure}
The structure under consideration is a lossless cylindrical cavity with radius $a$ and height $d$ as shown in FIG.~\ref{wedge}. The z-axis of the cavity is taken along the symmetry axis of the cavity. We investigate propagating solutions at angular frequency $\omega$ with a harmonic time dependence $e^{i\omega t}$ which is assumed and suppressed from all equations. 
For Transverse Electric (TE-to-z), the transverse components of the electromagnetic field can be determined from the axial component of the magnetic field $H_z$. The solution is
\begin{align}
H_z(\rho,\phi, z)=AJ_{\nu}\left (k_c\rho\right )\sin \left(\frac{p\pi z}{d}\right )e^{-i\nu \phi},\space  p=1,2,3
\label{axial_H}
\end{align}
Here, $A$ is an arbitrary constant, $\nu$ is the azimuthal propagation constant whose value is {\it\bf not known yet} and $J_{\nu}$ is the Bessel function of the first kind of order $\nu$. The parameter $k_c$ is related to the angular frequency by 
\begin{align}
    k_c^2+\left (\frac{p\pi}{d}\right )^2=\mu_o \epsilon_o \omega ^2=k_0^2 
    \label{Separation_TE}
\end{align}
The boundary condition at the perfectly conducting side walls of the cavity at $\rho =a$ is
\begin{align}
J_{\nu}'\left (k_ca \right)=0
\label{characteristic_TE}
\end{align}
It is well known that the only possible values of $\nu$ when $\phi$ covers the entire range $[0,2\pi]$ are integer values because of single-valuedness. It is important to emphasize at this point that we are not interested in solutions over the entire range of the azimuthal angle $\phi$. Instead, we are interested in  local solutions at a given value $\phi$. The situation is identical to how the propagation constant along a coaxial cable is determined: we enforce the boundary conditions only in a specified plane that is transverse to the direction of propagation. The equivalent in our case is to enforce the boundary conditions in a plane defined by a constant value of $\phi$. The characteristic equation is still Eq.~\ref{characteristic_TE} except that the value of $\nu$ is the unknown that we are trying to determine. The frequency is a parameter upon which the solution depends. In other words, the solution of this equation gives the dispersion relation $\nu=\nu(\omega)$. To highlight this point, equations Eq.~\ref{Separation_TE} and Eq.~\ref{characteristic_TE} are combined to yield
\begin{align}
J_{\nu}'\left(\sqrt{\mu_o\epsilon_o\omega^2a^2-\left (\frac{p\pi a}{d}\right )^2}\right )=0
\label{characteristic_TE_omega}
\end{align}
For a given value of the frequency $\omega$, assuming that the dimensions of the cavity are known, the only unknown in this equation is the azimuthal propagation constant $\nu=\nu(\omega)$. The dispersion relation can then be determined by sweeping over the frequency $\omega$ and determining the corresponding values of $\nu (\omega)$. Unfortunately, this method of solution is cumbersome since it requires a completely new solution for different dimensions of the cavity. Instead, a 'universal' solution consists in first determining the solutions of
\begin{align}
J_{\nu}'\left (x'\right )=0
\label{roots_TE}
\end{align}
and then combining these solutions with 
\begin{align}
\left |x'\right |=\sqrt{\mu\epsilon \omega^2a^2-\left (\frac{p\pi a}{d}\right )^2}.
\label{map_TE}
\end{align}
A simple method of solution consists in sweeping over $\nu$ starting from $\nu=0$ to determine the roots $x'$ from Eq.~\ref{roots_TE} and then use Eq.~\ref{map_TE} to relate $\nu$ to the angular frequency $\omega$. 
With the starting point set to $\nu=0$, Eq.~\ref{roots_TE} shows that there are an infinite number of solutions each corresponding to a solution of $J_0'(x')=-J_1(x')=0$. As $\nu$ is increased, a branch emerges from each root of $J_1(x')=0$. This implies that the lowest branch which determines the modes that are first to start to propagate in the azimuthal direction as the frequency is increased, is specified by the smallest root of $J_1(x')=0$ which happens to be $x'=0$. This root corresponds to $\nu=0$, which is an integer value for which the field is single-valued over the complete range of the azimuthal angle is in fact a resonance of the cavity.  In the literature on the subject, this solution is either ignored\cite{schelkunoff1943} or treated as trivial \cite{jackson2021classical}. The analysis in this paper shows that it is, in fact, the origin of the dominant branch. 
\begin{figure}[!h]
{\includegraphics[width=1\columnwidth]{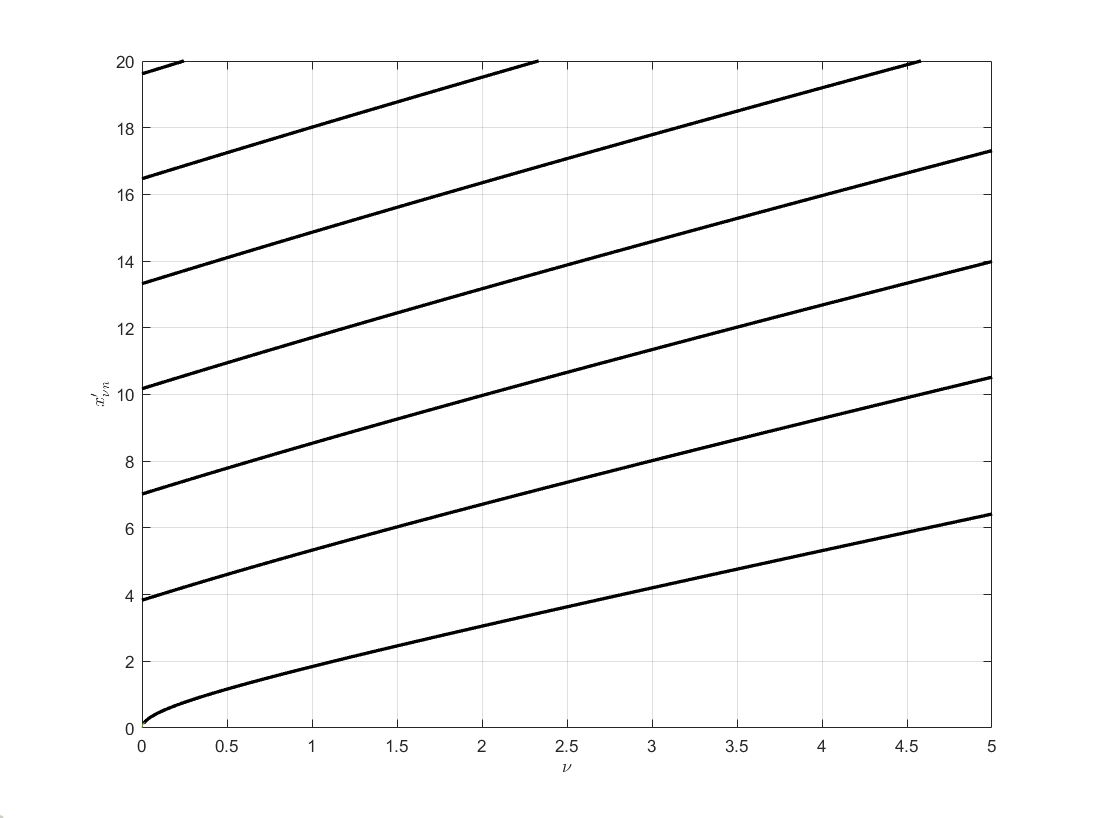}}
\centering
\caption{\label{TE_roots_versus_nu}First seven TE branches versus $\nu$. From each root of $J_1'(x')=0$ starts one branch.}
\end{figure}
Another interesting consequence of this study is the appearance of a cutoff frequency in the azimuthal direction that is different from that cutoff frequency int the axial direction. For real values of $\nu$, the relevant roots of $J_\nu '(x')=0$ are real and positive or zero. From Eq.~\ref{map_TE}, we then see that the first solution to start propagating in the azimuthal direction is given by using the root $x'=0$ in \ref{map_TE} leading to the condition
\begin{align}
\omega \geq \frac{p\pi}{d}\frac{1}{\sqrt{\mu_0\epsilon_0}}
\label{cutoff_TE}
\end{align}
Note that this cutoff frequency depends only on the height of the cavity and the material inside it. It does not depend on its radius. In particular, for an infinitely long cylinder, $d\to \infty$, propagation in the azimuthal direction can  take place at any non-zero value of the frequency. At $\omega=0$, the solution which corresponds to the resonance $(\nu,x')=(0,0)$ is too singular as will be shown later. For any non-zero value of $\omega$, waves can propagate in the azimuthal direction in the limit $d\to \infty$. However, these waves can not be observed as steady state solutions in an empty cylinder because of the requirement of single-valuedness which allows only integer values of $\nu$. 

The variation of the roots of Eq.~\ref{roots_TE} versus the azimuthal propagation constant $\nu$ is shown in FIG.~\ref{TE_roots_versus_nu}. Each branch starts from a root of Eq.~\ref{roots_TE} when $\nu=0$. Note that the lowest branch, which happens to contain the dominant mode of a cylindrical waveguide, i.e., the $TE_{11}$ mode starts from the 'resonance' at $\left (\nu, x'\right )=\left (0,0 \right )$. It is interesting to note that this resonance, that is rarely mentioned in the literature, appears as the 'origin' of the dominant azimuthal wave. 

To simplify the discussion, in what follows we assume that the z-dependence corresponds to $p=1$.
\subsection{Solutions of The Lowest TE Branch}
\begin{figure}[!h]
{\includegraphics[width=1\columnwidth]{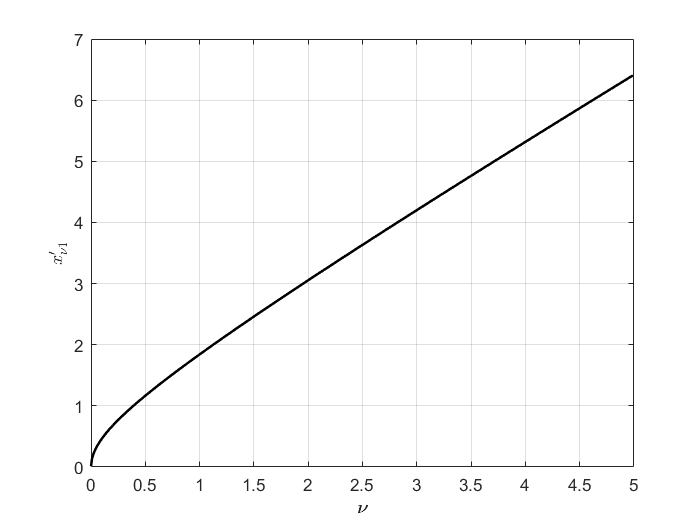}}
\centering
\caption{\label{lowest_TE_branch}Branch starting from smallest root $x_{\nu 1}'$ of $J_{\nu}'(x')=0$ versus $\nu$.}
\end{figure}

Let us denote by $x_{\nu 1}'$ the smallest (first) positive root of $J_{\nu}'(x')=0$ for a positive value of $\nu$. The lowest TE branch starts from the point $(\nu,x_{\nu 1}')=(0,0)$ as shown in Fig.~\ref{lowest_TE_branch} which gives the 'dispersion' relation $x_{\nu 1}'=x_{\nu 1}'\left (\nu\right )$. We see that  $x_{\nu 1}'$ is a monotonically increasing function of $\nu$ with an infinite slope at the origin.

\begin{figure}[!h]
{\includegraphics[width=1\columnwidth]{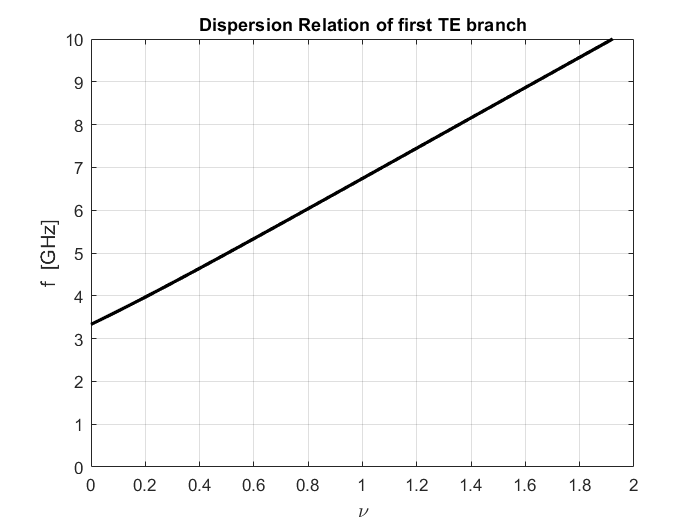}}
\centering
\caption{\label{lowest_TE_branch_dispersion}Dispersion relation $\nu = \nu\left ( f\right )$ of the lowest TE branch. $a= 15~$mm, $d=~45$mm.}
\end{figure}

The components of the electromagnetic field of the TE modes are given by ($p=1$ in the z-direction)
\begin{subequations}
\begin{align}
H_z(\rho,\phi, z,t)&=AJ_{\nu}(\frac{x_{\nu 1}'}{a}\rho)\sin(\frac{\pi z}{d})e^{-i\nu \phi}  \label{Hz}\\
H_\rho(\rho,\phi, z)&=A\frac{a\pi}{x_{\nu 1}'d}J_{\nu}'(\frac{x_{\nu 1}'}{a}\rho)\cos(\frac{\pi z}{d})e^{-i\nu \phi} \label{H_rho}\\
H_\phi(\rho,\phi, z)&=A\frac{i\pi\nu a^2}{\left (x_{\nu 1}'\right )^2d\rho}J_{\nu}(\frac{x_{\nu 1}'}{a}\rho)\cos(\frac{\pi z}{d})e^{-i\nu \phi} \label{H_phi}\\
E_\rho(\rho,\phi, z)&=-A\frac{\omega\mu\nu a^2}{\left ( x_{\nu 1}'\right )^2\rho} J_{\nu}(\frac{x_{\nu 1}'}{a}\rho)\sin(\frac{\pi z}{d})e^{-i\nu \phi}
\label{E_rho}\\
E_\phi(\rho,\phi, z)&=A\frac{i\omega \mu a}{x_{\nu 1}'}J_{\nu}'(\frac{x_{\nu 1}'}{a}\rho)\sin(\frac{\pi z}{d})e^{-i\nu\phi} \label{E_phi}
\end{align}
\end{subequations}
The angular frequency $\omega$ is related to the root $x_{\nu 1}'$ by 
\begin{align}
\left(\frac{x_{\nu 1}'}{a}\right )^2+\left(\frac{\pi}{d}\right )^2=\omega ^2\mu\epsilon 
\label{eq_separation_TE}
\end{align}
The dependence of the azimuthal propagation constant $\nu=\nu\left ( \omega\right )$ is obtained by combining the results shown in FIG.\ref{lowest_TE_branch} with Eq.~\ref{eq_separation_TE}. The results obtained for a radius $a=15$mm and a height of $d=45$mm are shown in FIG.\ref{lowest_TE_branch_dispersion}. We see that propagation in the azimuthal direction starts at f=10/3 GHz which is significantly lower than the cutoff frequency of the dominant mode of a cylinder of the same radius (5.86 GHz). It is also interesting to note that the dispersion relation of this mode can be accurately approximated by a linear function of frequency meaning that these waves are not dispersive, or equivalently, have a practically constant group velocity.  


The expressions of the components of the electromagnetic filed allow us to investigate their behaviour around the origin $(\rho=0)$. For simplicity, we keep only the radial variations of the different components and omit the azimuthal and axial variations which remain as given in above equations. Using the fact that $J_{\nu}\left ( x\right ) \simeq  \frac{1}{\Gamma \left ( \nu +1\right )}\left (\frac{x}{2}\right)^{\nu}$ and $J_{\nu}'\left ( x\right )\simeq  \frac{\nu}{x}\frac{\left ( x/2\right )^{\nu}}{\Gamma \left ( \nu +1\right )}$ for small values of $x$, we get the following local behaviour around the center of the cavity when $x_{\nu 1}'\rho/a \ll \sqrt {\nu+1}$
\begin{subequations}
\begin{align}
H_z &\simeq \frac{A}{\Gamma \left ( \nu +1\right )}\left ( x_{\nu 1}'\frac{\rho}{a}\right )^{\nu}\\
H_{\rho}&\simeq \frac{A a\pi}{\Gamma \left ( \nu +1\right )d}\frac{\nu}{\left (x_{\nu 1}'\right )^2}\left ( \frac{x_{\nu 1}'}{2}\right )^{\nu}\left (\frac{\rho}{a} \right )^{\nu-1}\\
H_\phi &\simeq -\frac{iA\pi a}{\Gamma (\nu +1)d}\frac{\nu }{\left (x_{\nu 1}'\right )^2}\left (\frac{x_{\nu 1}'}{2}\right )^{\nu}\left ( \frac{\rho}{a}\right )^{\nu -1}\\
E_{\rho}&\simeq -\frac{A\omega \mu a}{\Gamma (\nu +1)}\frac{\nu }{\left (x_{\nu 1}'\right )^2}\left (\frac{x_{\nu 1}'}{2}\right )^{\nu}\left ( \frac{\rho}{a}\right )^{\nu -1}\\
E_{\phi}&\simeq -\frac{iA\omega \mu a}{\Gamma (\nu +1)}\frac{\nu }{\left (x_{\nu 1}'\right )^2}\left (\frac{x_{\nu 1}'}{2}\right )^{\nu}\left ( \frac{\rho}{a}\right )^{\nu -1}
\label{local_TE}
\end{align}
\end{subequations}
The equations clearly show that the transverse components of the electromagnetic field are singular for values of $0<\nu < 1$. The smaller the value of $\nu$ the stronger the singularity at $\rho =0$. The axial component $H_z$ remains finite. Note that although some components of the electromagnetic field are singular around the origin, the energy stored in the cavity remains finite for $\nu \neq 0$. There is no physical reason for this type of waves not to propagate in the structure since they satisfy Maxwell's equations with the boundary conditions and have finite energy.
\subsection{Lowest Branch Solution for Small Values of $\nu$}
As stated earlier, the pair $\left (\nu,x_{\nu 1}'\right )=\left (0,0\right )$ is a solution to the characteristic equation \ref{roots_TE}. It is, however, important to know the behaviour of the electromagnetic field of this solution. From the equations \ref{local_TE}, it is seen that the transverse components contain the terms $\frac{\nu}{\left (x_{\nu 1}'\right )^2}$ and $\left (\frac{x_{\nu 1}'}{2}\right )^{\nu} $. Since both $\nu$ and $x_{\nu 1}'$ approach zero, these terms require a more careful analysis. 

From the small argument approximation of the Bessel functions, Eq.~\ref{roots_TE} becomes
\begin{align}
    J_{\nu}'(x')&=\frac{\nu J_{\nu}\left (x'\right )}{x}-J_{\nu +1}\left ( x'\right )\\ \nonumber
    &\simeq \frac{\nu}{x'\Gamma\left (\nu +1\right )}\left (\frac{x'}{2}\right )^{\nu}-\frac{1}{\Gamma \left (\nu +2\right )}\left ( \frac{x'}{2}\right )^{\nu +1}\\ \nonumber
    &=\frac{1}{\Gamma (\nu+1)}\left (\frac{x'}{2} \right )^{\nu}\left [ \frac{\nu}{x'}-\frac{\Gamma(\nu+1)}{\Gamma (\nu+2)}\frac{x'}{2}\right ]=0
\end{align}
From this equation, using the property of the $\Gamma$ function $\Gamma \left (\nu+1\right )=\nu \Gamma \left (\nu \right )$, we obtain
\begin{align}
 x'\approx \sqrt{2\nu},\qquad \nu \to 0  
\end{align}
With this result, we see that the two problematic terms are both finite with
\begin{align} \nonumber
\lim _{\nu \to 0} \frac{\nu}{\left (x_{\nu 1}'\right )^2}=\frac{1}{2}\\
\lim _{\nu \to 0}\left (\frac{x_{\nu 1}'}{2}\right )^{\nu}=1 \nonumber
\end{align}
With these results, we get the expressions of the electromagnetic field components of the lowest TE branch when $\nu \to 0$
\begin{subequations}
\begin{align}
H_z &\simeq \frac{A}{\Gamma \left ( \nu +1\right )}\left (\frac{\rho}{a}\right )^{\nu}\\
H_{\rho}&\simeq \frac{A a\pi}{2\Gamma \left ( \nu +1\right )d}\left (\frac{\rho}{a} \right )^{\nu-1}\\
H_\phi &\simeq -\frac{iA\pi a}{2\Gamma (\nu +1)d}\left ( \frac{\rho}{a}\right )^{\nu -1}\\
E_{\rho}&\simeq -\frac{A\omega \mu a}{2\Gamma (\nu +1)}\left ( \frac{\rho}{a}\right )^{\nu -1}\\
E_{\phi}&\simeq -\frac{iA\omega \mu a}{2\Gamma (\nu +1)}\left ( \frac{\rho}{a}\right )^{\nu -1}
\label{small_nu_TE}
\end{align}
\end{subequations}
The transverse components of the electromagnetic are all singular at the center where they behave as $\rho ^{\nu -1}$. The energy in the solution remains finite for all non-zero values of $\nu$. If we take the solution at $\left (\nu,x_{\nu 1}' \right )=\left (0,0 \right )$ as the limit when $\nu \to 0$, we see that the singularity becomes $\rho ^{-1}$. This solution has an infinite energy and can not be excited. However, the energy remains finite for any non-zero value of $\nu$. One can only wonder whether it can exist in lossy systems with tremendous energy resources and whether it is involved in cosmic jets for example\cite{10.1093/pasj/54.3.L39}. 
\section{Transverse Magnetic Modes}
The Transverse Magnetic (TM) modes can be analyzed similarly to the TE modes. The components of the electromagnetic field can be obtained from the axial component of the electric field, $E_z$. The resulting expressions are $\left ( H_z=0\right )$
\begin{subequations}
\begin{align}
E_{\rho}&=A\frac{p\pi a}{x_{\nu n}d}J_{\nu}'\left (\frac{x_{\nu n}}{a}\rho \right )e^{-i\nu \phi}\sin \left (\frac{p\pi z}{d}\right )\\
E_{\phi}&=-A\frac{i\nu a^2}{x_{\nu n}^2\rho}\frac{p\pi}{d}J_{\nu}\left (\frac{x_{\nu n}}{a}\rho \right )e^{-i\nu \phi}\sin \left (\frac{p\pi z}{d}\right )\\
E_z&=AJ_{\nu}\left (\frac{x_{\nu n}}{a}\rho \right )e^{-i\nu \phi}\cos \left (\frac{p\pi z}{d}\right )\\
H_{\rho}&=A\frac{\omega \epsilon \nu a^2}{x_{\nu n}^2\rho }J_{\nu}\left (\frac{x_{\nu n}}{a}\rho \right )e^{-i\nu \phi}\cos \left (\frac{p\pi z}{d}\right )\\
H_{\phi}&=-A\frac{i\omega \epsilon a}{x_{\nu n} }J_{\nu}'\left (\frac{x_{\nu n}}{a}\rho \right )e^{-i\nu \phi}\cos \left (\frac{p\pi z}{d}\right )
\end{align}
\end{subequations}
Here, $p=0,1,\dots$, $x_{\nu n},\quad n=0,1,\dots$ is the n$^{th}$ root of $J_{\nu}\left ( x\right )$ and $A$ is an arbitrary constant.
\begin{figure}[!h]
{\includegraphics[width=1\columnwidth]{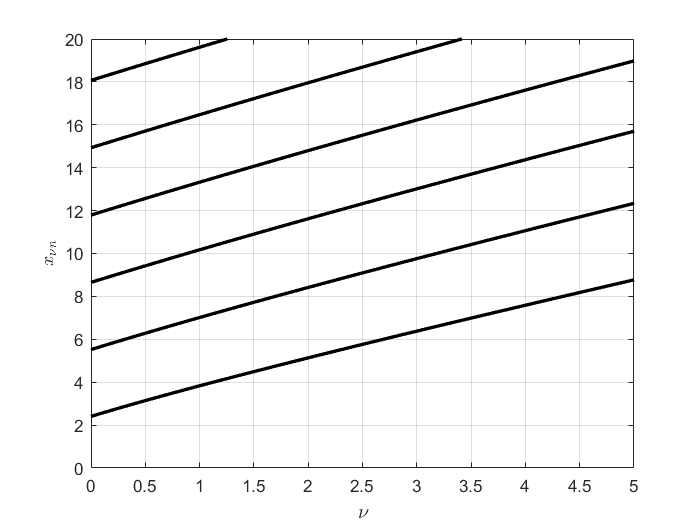}}
\centering
\caption{\label{all_branches_TM} Variation of $x_{nu n},n=1,\dots, 6$ for the first six TM modes. $x_{\nu 1}'$ of $J_{\nu}'(x')=0$ versus $\nu$.}
\end{figure}
Following the analysis of the TE modes, the dispersion relation of the TM modes is obtained from the two-step process. We first determine the roots of the characteristic equation
\begin{align}
  J_{\nu}\left (x\right )=0.
  \label{roots_TM}
\end{align}
The relationship between the azimuthal propagation constant $\nu$ and the frequency is then obtained from Eq.~\ref{map_TE} in which the root $x'$ is replaced by the root given by Eq.~\ref{roots_TM}. The solutions are again grouped in branches with each branch starting from a root of $J_0\left ( x\right )=0$. The lowest branch starts from the solution $x=2.40483$. The characteristic equation \ref{roots_TM} admits the solution $x=0$ except for $\nu=0$. However, it can be shown from the expressions of the components of the electromagnetic field given above that this solution corresponds to the trivial solution $A=0$ even when $\nu$ is not an integer. 

For integer values of $\nu$, which is the case for a steady state solution where the full range $0\leq \phi \leq 2\pi$ is accessible, the components of the EM field vanish or remain finite at the center of cylinder $\rho =0$. However, this is not the case when $0<\nu <1$. Indeed, the transverse components of the electromagnetic field behave as $\rho ^{\nu -1}$ in the vicinity of $\rho =0$. More specifically, for $x_{\nu n}\rho /a \ll \sqrt{\nu +1}$ we  have the local approximations (the axial and azimuthal dependence is suppressed for simplicity)
\begin{subequations}
\begin{align}
E_z &\simeq \frac{A}{\Gamma \left ( \nu +1\right )}\left (\frac{x_{\nu n}}{2}\right )\left ( x_{\nu 1}\frac{\rho}{a}\right )^{\nu}\\
E_{\rho}&\simeq \frac{A ap\pi}{\Gamma \left ( \nu +1\right )d}\frac{\nu}{\left (x_{\nu n}\right )^2}\left ( \frac{x_{\nu n}}{2}\right )^{\nu}\left (\frac{\rho}{a} \right )^{\nu-1}\\
E_\phi &\simeq -\frac{iAp\pi a}{\Gamma (\nu +1)d}\frac{\nu }{\left (x_{\nu n}\right )^2}\left (\frac{x_{\nu n}}{2}\right )^{\nu}\left ( \frac{\rho}{a}\right )^{\nu -1}\\
H_{\rho}&\simeq -\frac{A\omega \epsilon a}{\Gamma (\nu +1)}\frac{\nu }{\left (x_{\nu n}\right )^2}\left (\frac{x_{\nu n}}{2}\right )^{\nu}\left ( \frac{\rho}{a}\right )^{\nu -1}\\
H_{\phi}&\simeq -\frac{iA\omega \epsilon a}{\Gamma (\nu +1)}\frac{\nu }{\left (x_{\nu n}\right )^2}\left (\frac{x_{\nu n}}{2}\right )^{\nu}\left ( \frac{\rho}{a}\right )^{\nu -1}
\end{align}
\end{subequations}
As in the case of the TE modes, we see that the transverse components of the electromagnetic field are singular at the center of the cavity when $0<\nu <1$. There is, however, a substantial difference between the TE and TM modes especially in regards to the region of space where the local approximations are valid. The small argument approximation of the Bessel function is valid for $\rho /a \ll \frac{\sqrt{\nu +1}}{x_{\nu n}}$. For TM modes, the smallest root is $x_{01}=2.40483$. This means that the expression of the transverse field components of the lowest TM mode around the center of the cavity are valid only when $\rho/a\ll 1/x_{01}\approx 0.5$. On the other hand, the smallest root for the TE modes is $x'=0$. In this limit, the transverse components of the lowest TE mode behave as $\rho ^{-1}$ over the entire cross section of the cavity. In addition, solutions that correspond to larger values of $x_{\nu n}$ or $x_{\nu n}'$ are singular only over an increasingly smaller region centered at $\rho=0$ and increasingly higher frequencies. For integer values of $\nu$, the field components are all finite at the center of the cavity for $\nu =0$ except for the lowest TE branch. 

\section{Observation of Singular Solutions}
As we have seen, the azimuthally propagating solutions are singular at the center of the cavity for $\nu <1$. Obviously, these fields can be exist as steady sate solution in a uniform cavity where $\nu$ is restricted to integer values.
  
One way to excite the singular modes is to impose boundary conditions in the azimuthal direction that result in resonances with $\nu <1$. This can be achieved by inserting two perfectly conducting surfaces at form a metal wedge with internal angle $\phi$. The electromagnetic field in the space outside the metallic wedge is a superposition of two azimuthal waves propagating in the $+\phi$ and $-\phi$ directions. For example, for a TM mode,the axial electric field  takes the form
  \begin{align}
      E_z=J_{\nu}\left (\frac{x_{\nu n}}{a}\right )\left [ Ae^{-i\nu \phi}+Be^{i\nu \phi} \right ]\cos \left (\frac{\pi z}{d}\right )
      \label{TM_wedge}
  \end{align}
  If the faces of the metallic wedge are located at $\phi=\theta$ and $\phi=2\pi-\theta$, the vanishing of $E_z$ at these planes leads to the possible values of $\nu$
  \begin{align}
      \nu = \frac{m\pi}{2\pi-\theta},\quad m=1,2,\dots
  \end{align}
  The smallest value of $\nu$ results when $m=1$, or
  \begin{align}
      v_{min}=\frac{\pi}{2\pi -\theta}
      \label{nu_min}
  \end{align}
  As $\theta$ varies from $0$ to $\pi$, $\nu _{min}$ varies from 0.5 to 1 thereby providing a mechanism to observe some of the singular solutions.
  The case of TE modes can be handled similarly. It turns out that the expression of the azimuthal propagation constant 0f the TE modes is also given by Eq.~\ref{nu_min} except that $\nu=0$ is a solution in this case.
  Once the values of $\nu$ are known, the resonant frequency can be determined directed from FIG.~\ref{lowest_TE_branch_dispersion} without any further calculations. The results of the lowest TE resonance are shown in FIG.6. For comparison, results from HFSS are shown. The agreement between the two is excellent.
    
\begin{figure}[!h]
{\includegraphics[width=1\columnwidth]{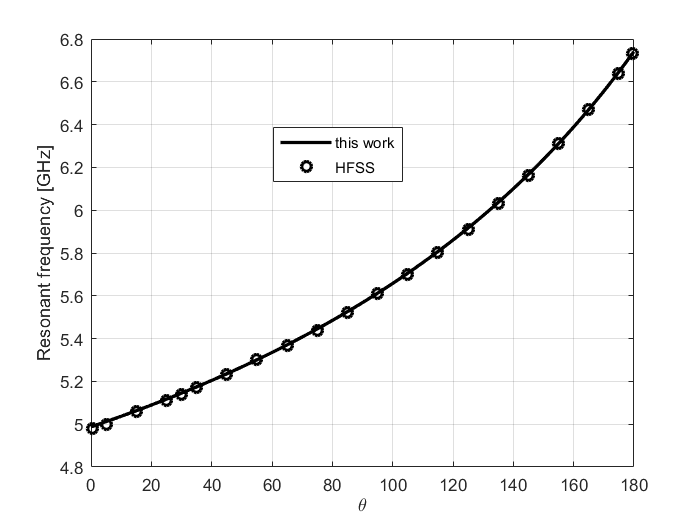}}
\centering
\caption{\label{Wedge_Resonance} Resonant frequency of the lowest TE mode of a cavity with a metallic wedge versus internal angle $\theta$ for $a= ~$mm, $d=~$mm. The circles are from the commercial software package HFSS.}
\end{figure}

To force resonances for values of $\nu<0.5$ artificial surfaces provide a possible route. Indeed, if one of the faces of the wedge is assumed to be a perfect magnetic wall and the other a perfect electric wall, the minimum value of $\nu$ is easily shown to be $\nu _{min}=0.25$. For further decrease in the value of $\nu$ below 0.25 we could consider capacitively loading of one of the surfaces similarly to the technique used to reduce the size of resonant antennas. Numerical simulations using HFSS show that this is indeed possible. However, the strong singularity of the transverse fields for small values of $\nu$, say around 0.1, is numerically very demanding, especially for a method such the finite element method upon which HFSS is based. 

It should be mentioned here that singular fields in the vicinity of sharp metallic wedge has been known for a long time \cite{collin1991guided}. However, this local feature was viewed as a consequence of the boundary conditions. This study shows that they are in fact properties of the azimuthally propagating waves. The boundary conditions force these waves to resonate and exhibit their intrinsic singularity.

An alternative way to observe these waves for non-integer values of $\nu$ is to excite transient waves. Currents that support these waves can be generated by moving electrically charged fluids with their motion restricted to only part of the range $[0,2\pi]$ and over short time intervals. Such signals would have broad bandwidth and couple to solutions with small values of $\nu$. Whether the initiation of lightening, which is still not understood\cite{BUITINK20101}, can be explained by these transients is a very interesting question that we are currently investigating.

\section{Application to Dual-Mode RF Filters}
The original goal of this study was to develop models of microwave filters in circular cavities based on propagation and not resonance. To that end, the variation of the propagation constant of azimuthally propagating modes as a function of frequency is required. As argued in earlier sections, when applicable, propagation-based models contain significantly more information than resonance-based models. Fortunately, robust microwave filter designs can be carried out using cavities in which modes can propagate in uniform sections that are separated by discontinuities. We give a simple illustrative example here; more complex designs will be presented in future articles. Even though the appearance of the lowest branch in the TE mode is arguably the most surprising result of this study, actual applications involve mainly the part of the branch that is located around $\nu=1$. In that sense, the fact that solutions around $\nu=0$ do not seem to be accessible, is of little important to practical applications such as dual-mode filters. 
\begin{figure}[!h]
{\includegraphics[width=1\columnwidth]{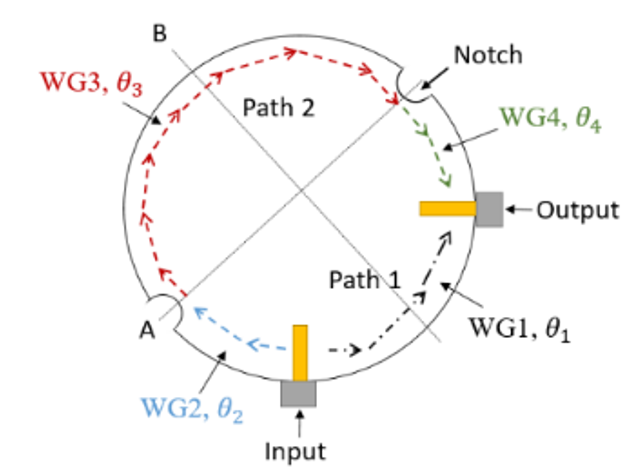}}
\centering
\caption{\label{pert_cavity} Top view of a circular waveguide cavity with coaxial input/output and longitudinal grooves.}
\end{figure}
\subsection{Design Example}

\begin{figure}[!t]
{\includegraphics[width=1\columnwidth]{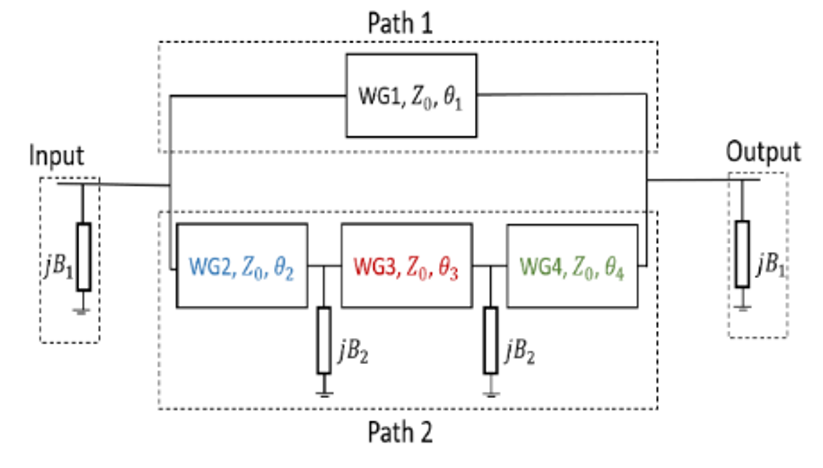}}
\centering
\caption{\label{circuit} Equivalent circuit based on propagating waves in the azimuthal direction.}
\end{figure}

We consider a dual-mode circular cavity as shown in FIG.~\ref{pert_cavity}. The input and output are coaxial probes that are placed 90$^{\circ}$ apart. Perturbations that extend over the total height of the cavity are placed midway between the probes at 135$^{\circ}$. The radius and the height of the cavity are $a=$15 mm and $d=$45 mm, respectively. With these dimensions, the resonant frequency of the degenerate TE$_{111}$ can be read directly from FIG.\ref{lowest_TE_branch_dispersion}  resulting in fr = 6.74 GHz. Incoming energy from the input to the output of the dual-mode cavity splits into two azimuthally propagating waves in the clockwise and counterclockwise directions. The dimensions of the cavity are such that the TE mode of the lowest branch is propagating in the azimuthal direction around $\nu=1$ at the center of the passband of the filter. Destructive interference between these waves at the output is be used to generate transmission zeros (TZ)’s at finite frequencies. The equivalent model shown in FIG.~\ref{circuit}  is based on azimuthally propagating waves in opposing directions. These are characterised by their 'electric length $\nu \theta$, the characteristic impedance $Z_0$ which is set to unity. The electric length is related to the angular length by using a linear approximation of the dispersion relation in FIG.\ref{lowest_TE_branch_dispersion} around $\nu =1$.
\begin{align}
\theta^{\circ} = \theta^{r} \frac{f + f_{rTE211} - 2\times f_{rTE111}}{f_{rTE211} - f_{rTE111}}
\end{align}
\begin{figure}[!t]
{\includegraphics[width=1\columnwidth]{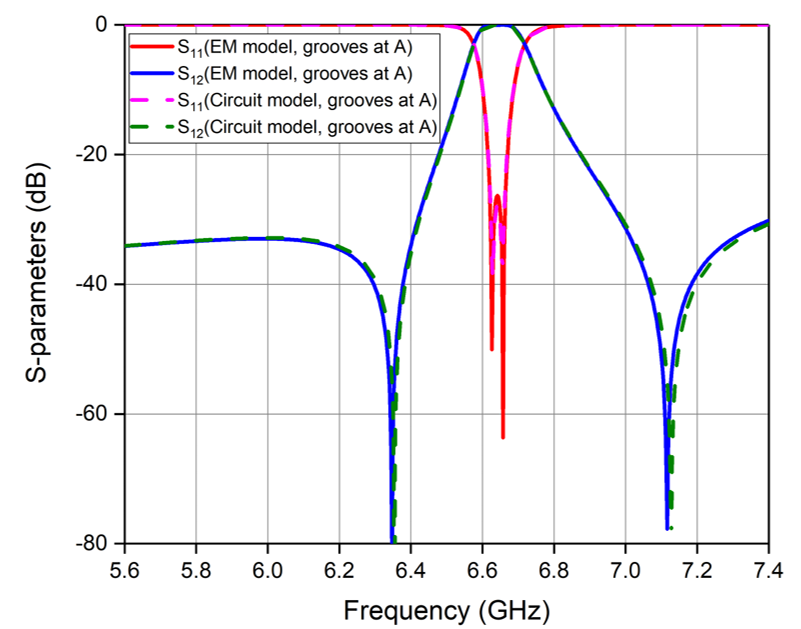}}
\centering
\caption{\label{grovA} Comparison between the scattering response of the propagation-based mode and EM full-wave simulation with grooves as location A.}
\end{figure}

\begin{figure}[!t]
{\includegraphics[width=1\columnwidth]{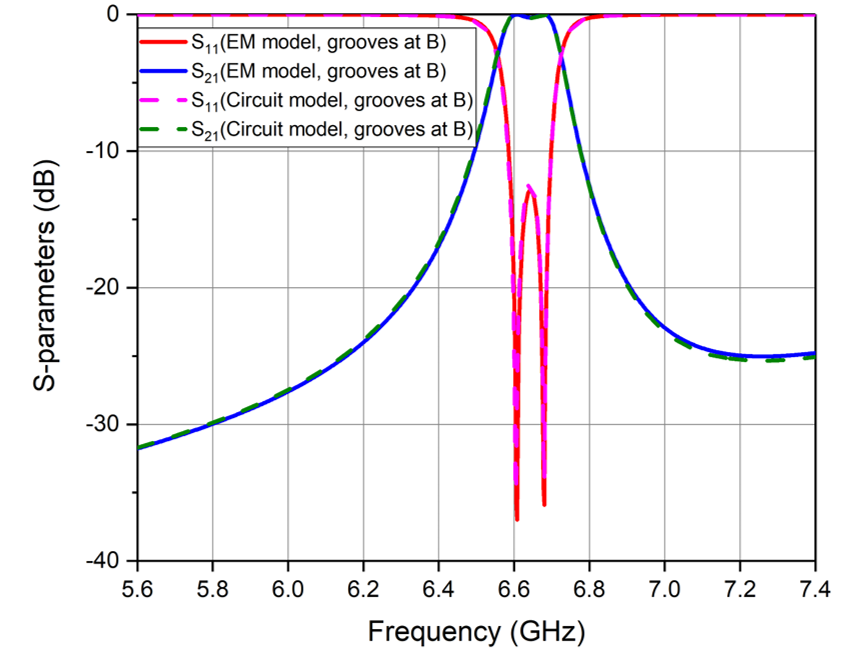}}
\centering
\caption{\label{grovB} Comparison between the scattering response of the propagation-based mode and EM full-wave simulation with grooves as location B.}
\end{figure}
Here, f$_{rTE111}$ and f$_{rTE211}$ are the resonant frequencies of the TE$_{211}$ and TE$_{111}$. The quantity $\theta^{\circ}$ is the electric angular length and $\theta^{r}$ the angular length. The approximation is valid for the lowest branch in the TE mode around $\nu=1$. For narrowband applications, the frequency dependence of the wave impedance is negligible and will be ignored in this example. The normalised circuit parameters of the branch line model in Fig are Z$_{\circ}$ = 1, B$_{1}$ = 0.0242, B$_{2}$ = 0.0096. The parameter B$_{1}$ is preceded and followed by a transformer at the input and the  output with normalised impedance of 4.9751 Ohm. A comparison between the response of the equivalent circuit and the HFSS model is shown in FIG.~\ref{grovA} \& FIG.~\ref{grovB}. 

\begin{figure}[!h]
{\includegraphics[width=1\columnwidth]{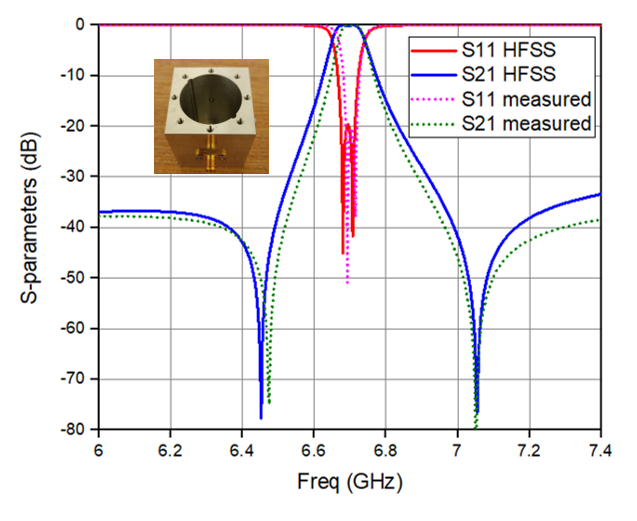}}
\centering
\caption{\label{meas} Comparison between the scattering response of the EM model and the measured hardware.}
\end{figure}

The solid lines correspond to the HFSS model, and the dotted lines correspond to the propagation-based model. The agreement is excellent. The model accurately predicts the appearance of a second order bandpass filter and two real frequency transmission zeros due to constructive and destructive interference of the two waves propagating in opposing azimuthal directions. The model goes further to show the dependence of transmission zeros on the nature and location of perturbations. Here, the grooves are of inductive nature. As seen in FIG.~\ref{grovA} , the two transmission zeros are located on the imaginary axis of the s-plane when the grooves are at location A whereas the transmission zeros become complex when the grooves are at location B as in FIG.~\ref{grovB}. The dispersion of the coupling elements and perturbations cause the weak asymmetry in the frequency response. The structure can be cascaded to obtain higher-order filters having arbitrary N real-frequency transmission zeros. The design methodology and physical dimensioning will be detailed in future papers. A prototype hardware has been fabricated and measured as shown in FIG.~\ref{meas}, demonstrating the validity of the detailed analysis in this paper.



\bibliographystyle{naturemag}

\bibliography{refs}

\end{document}